\documentstyle[epsf]{elsart}  
\begin{document}
\newcommand{\p}{\partial}
\newcommand{\ls}{\left(}
\newcommand{\rs}{\right)}
\newcommand{\beq}{\begin{equation}}
\newcommand{\eeq}{\end{equation}}
\newcommand{\beqa}{\begin{eqnarray}}
\newcommand{\eeqa}{\end{eqnarray}}
\newcommand{\bdm}{\begin{displaymath}}
\newcommand{\edm}{\end{displaymath}}
\begin{frontmatter}
\title{In-medium dependence and Coulomb effects of the 
pion production in heavy ion collisions} 
\author{V.S. Uma Maheswari, C. Fuchs, Amand Faessler,  L. Sehn, 
D.S. Kosov, Z. Wang}
\address{Institut f\"ur Theoretische Physik der 
Universit\"at T\"ubingen, D-72076 T\"ubingen, Germany}
\begin{abstract}
The properties of the high energy pions observed in heavy ion 
collisions, in particular in the system Au on Au at 1 GeV/nucleon 
are investigated. 
The reaction dynamics is described within the Quantum Molecular Dynamics (QMD)
approach. It is shown that high energy pions freeze out early and originate
from the hot, compressed matter. $N^*$--resonances are found to give 
an importnat contribution toward the high energy tail of the 
pion. Further the role of in-medium effects in the description of charged pion
yield and spectra is investigated using a microscopic 
potential derived from the Brueckner G-matrix which is 
obtained with the Reid soft-core potential.
It is seen that the high energy part of the spectra is relatively
more suppressed due to in-medium effects as compared to the
low energy part. A comparision to
experiments further demonstrates that the present calculations
describe reasonably well the neutral (TAPS) and charged (FOPI) pion
spectra. The observed energy dependence of the $\pi^-/\pi^+$ ratio, 
i.e. deviations from the isobar model prediction, is due to Coulomb effects 
and again indicate that high energy
pions probe the hot and dense phase of the reaction. These 
findings are confirmed independently by a simple phase space analysis. 
\end{abstract}
\begin{keyword}
Pion spectra, isospin dependence, $\pi^-/\pi^+$ ratios, QMD
\\
PACS numbers: {\bf 25.75.+r}
\end{keyword}
\end{frontmatter}
\section{Introduction}
Heavy ion collisions play a vital role in the quest for
the nuclear equation of state,
since a hot and dense nuclear matter phase is created, 
though transiently, in heavy ion collisions.
In a typical intermediate energy reaction, for e.g. 
Au+Au at 1 GeV/A, 
matter gets compressed to densities of about two to three times 
the normal nuclear matter density $\rho_0$ and the 
maximum temperature attained 
in this central zone is supposed to reach values of 
few tens to about 100 MeV \cite{puri92,maru94,bass95}.

The investigation of particle production is a well established approach
to probe the nuclear matter under these extreme 
conditions \cite{naga81,broc84,stoc86}.
At incident energies around 1 GeV/A the
pion production is the dominant channel
by which the hot, compressed matter de-excites.
Pion spectra at these relativistic energies 
have been measured at GSI (FOPI \cite{pelt96}, KaoS \cite{bril93,munt95}, 
TAPS \cite{vene93,schw94}) and LBL \cite{naga81,broc84}. Unlike at 
lower energies the spectra show non-thermal 
slopes, i.e. one needs a superposition of
two Maxwell-Boltzmann distributions to describe
the data \cite{broc84,pelt96,munt95,schw94,baur91,brau95}.
 The onset of the second slope roughly
coincides with the threshold for the $\Delta-$resonance. 
Therefore, the resonance production
and its subsequent decay into pions is considered
to be important for an understanding of the 
excited nuclear matter.
For this reason, in addition to the $\Delta-$channel
we include one-pion and two-pion decays of
$N^*$--resonances and the respective one-pion reabsorption process.
The role of these resonances in the production of
the high energy pions is analysed in this work.
Furthermore, recent measurements indicate that the high energy pions
freeze out early, particularly when emitted perpendicular
to the reaction plane \cite{bril93,bass93}. 
Therefore, these pions most likely
probe the early phase of the collision.
Hence, a study of 
subthreshold pions is of current interest and is
being actively pursued by both experimentalists
and theoreticians.

As compared to the nature of subthreshold pions, 
the properties of the low energy pions, which are
interpreted to be originating from $\Delta-$decay,
 are relatively well understood within transport
model approaches. However, one finds that the
transport models such as QMD and BUU, in general, 
overestimate the charged pion yields as well as
the low energy part of the spectra 
\cite{pelt96,teis96}.
In this context, one has to keep in mind that 
the nuclear mean field, in particular, the
respective momentum dependence of the
interaction has large effects on various 
dynamical observables \cite{jaen92,khoa92,gale87}. 
In most of the previous studies on pion
production a nuclear mean field derived from 
effective phenomenological interactions, 
such as the Skyrme forces, have
been applied \cite{bass95,pelt96,teis96,teis97,bass94}.
However, a more realistic description 
of baryon dynamics in the hot and compressed phase
of nuclear matter is possible within the framework
of Brueckner theory where the nuclear mean field 
is derived microscopically from the G-matrix
obtained by solving either the Bethe-Goldstone 
or the Bethe-Salpeter equation.
With these realistic forces one finds in general 
an enhancement of the repulsive part of the 
nuclear interaction which originates from the
strong momentum dependence of the G-matrix 
\cite{jaen92,khoa92,li93,sehn97}.
Since observables such as nuclear flow
are found to be sensitive to the use
of realistic forces \cite{khoa92,rami95,fuchs96a}, 
it is a worthwhile task to analyse the role of in-medium effects in the
description of pionic observables at SIS energies.
Therefore, in the present work, we also investigate
the effect of the Brueckner G-matrix potential on
pion multiplicities and spectra and then compare to
the results obtained with the standard Skyrme interaction.

Another interesting aspect of the pion production 
observed in heavy ion collisions is its charge
dependence \cite{bene79,libh79,bert80,gyul81,osad96}. 
In the case of heavy systems such as 
Au one finds \cite{pelt96,munt95} 
for the total abundancies as a general law
 $\sigma (\pi^- ) \ > 
\sigma (\pi^0 ) > \sigma (\pi^+ ) $. The two 
principal effects contributing to the observed
charge dependence are the isopin dependence of the pion
production cross sections and
the Coulomb force. Since most of the pions
observed at bombarding energies around 1 GeV/A originate from
the decay of $\Delta-$resonances, one can determine
the isospin effect in a simple isobar model \cite{stoc86,west82,eric88},
 which depends
only on the $Z/A$ ratio of the colliding systems and is
independent of the pion energy. 
However, in Refs. \cite{pelt96,munt95} deviation
from the behaviour predicted by the isobar model have been
observed. The observed energy dependence of the $\pi^- / \pi^+ $ ratio is
then essentially due to the other factor, i.e. the Coulomb effect 
which modifies the pion momentum and 
the available phase space \cite{gyul81}. 
As mentioned in Ref. \cite{oesc96} 
one can moreover use the energy dependence of this ratio 
to estimate the size of the
pion emitting source and thereby the density of the
participant zone probed by the observed pions. This
provides us with an alternative approach to investigate
the nature of the subthreshold pions.
To study this aspect, we take into account the Coulomb
interaction between the baryons and the
 charged pions. This contributes toward the real part
of the pion optical potential. Another contribution arises
from the effect of nuclear medium on the pion dispersion relation, 
i.e. from collective $\Delta N^{-1}$ and $NN^{-1}$
excitations. However, it is shown in a recent study \cite{fuch96} that 
only weak corrections, i.e. a small in-medium pion-nucleus
potential, are required in order to obtain a realistic description
of pionic observables. In addition, also in other 
works \cite{eheh93,xion93} it was found that the high energy 
part of the pion spectrum is not much affected by these
medium corrections. 
A more sophisticated treatment of collective pionic 
excitations within the transport approach was proposed
in Ref.\cite{helg95} which was subsequently applied to
heavy ion collisions in Ref.\cite{helg97}. However, as
discussed in Ref.\cite{fuchs97}, it is rather questionable
if the simple $\Delta N^{-1}$ ( and $NN^{-1}$)
model yields a realiable pion dispersion relation.
Further, the results of Ref.\cite{helg97} indicate
that the influence of collective excitations on
pionic observables is small and tends to enhance the
pion yields with respect to the standard approaches 
and the experimental findings.
Hence, we omit such colective medium effects on 
the charge dependence of pion production in the present work.

Here we mainly focus upon the aforesaid aspects and
investigate them using the Quantum Molecular Dynamics (QMD) 
model \cite{Ai91}. The essentials of the QMD model is 
briefly introduced in section 2. Results 
obtained are discussed in detail in section 3. Finally, we summarize
our findings in section 4.
\section{QMD model }
Quantum molecular dynamics 
is a semiclassical transport model which accounts for relevant 
quantum aspects like the Fermi motion of the nucleons, stochastic 
scattering processes including Pauli blocking in the final states, 
the creation and reabsorption of resonances
and the particle production. A detailed description 
of the QMD approach can be found in Refs. \cite{khoa92,Ai91}.

The time evolution of the individual nucleons 
is thereby governed by the classical equations of motion
\beqa
\frac{\p {\vec p}_i}{\p t} = - \frac{\p H}{\p {\vec q}_i}\quad ,\quad 
\frac{\p {\vec q}_i}{\p t} =   \frac{\p H }{\p {\vec p}_i}\quad  
\label{motion1}
\eeqa
with the classical $N-$body Hamiltonian $H$
\beq
H = \sum_i \sqrt{{\vec p}_{i}^2 + M_{i}^2} + 
\half \sum_{i,j\atop (j\neq i)} \left( U_{ij} + U_{ij}^{\rm Yuk} 
+ U_{ij}^{\rm Coul}\right)
\quad .
\label{hamB}
\eeq
The Hamiltonian, Eq. (\ref{hamB}), contains mutual two- (and three-) 
body potential interactions which are finally determined as 
classical expectation values
from local Skyrme forces $ U_{ij}$ supplemented by a phenomenological
momentum dependence and an effective 
Coulomb interaction $U_{ij}^{\rm Coul}$, 
\beqa
 U_{ij} &=& \alpha\left(\frac{\rho_{ij}}{\rho_0}\right)
+\beta\left(\frac{\rho_{ij}}{\rho_0}\right)^{\gamma}
+\delta ln^2\left(\epsilon |{\vec p}_i -{\vec p}_j |^2 +1 \right)
\frac{\rho_{ij}}{\rho_0},
\label{skyrme} 
\\
 U_{ij}^{coul} &=& {\left ( \frac{Z}{A}\right )}^2 
\frac{e^2}{|{\vec q}_i- {\vec q}_j |} 
erf\left(\frac{|{\vec q}_i-{\vec q}_j|}{\sqrt{4L}}\right),
\label{coul}
\eeqa 
where $\rho_{ij}$ is a two-body interaction density defined as 
\beq
\rho_{ij} = \frac{1}{(4\pi L)^{\frac{3}{2}}}
\e^{-( {\vec q}_{i}- {\vec q}_{j})^2 /4L},
\label{dens1}
\eeq
and $erf$ is the error function. The parameters $\alpha, \beta, \gamma, 
\delta, \epsilon$ of the Skyrme interaction, Eq. (\ref{skyrme}), 
are determined in order to
reproduce simultanously the saturation density 
($\rho_0 = 0.16 fm^{-3}$) and the binding energy ($E_B=-16$ MeV) for
normal nuclear matter for a given incompressibility $K_{\infty }$
 as well as the correct
momentum dependence of the real part of the 
nucleon-nucleus optical potential \cite{khoa92,Ai86}. In the present 
calculations we use a soft equation of state 
with $K_{\infty }=$ 200 MeV including the momentum 
dependent interaction (SMD), Eq. (\ref{skyrme}). 
 The Yukawa-type potential 
$ U_{ij}^{\rm Yuk}$ in Eq. (\ref{hamB}) mainly serves to improve 
the surface properties and the stability of the initialized nuclei. 

Nucleons while propagating shall collide stochastically, if the
distance between the centroids of the two Gaussian wavepackets 
is less than $d_{\rm min}=\sqrt {\sigma_{\rm tot}({\sqrt s})/\pi }$.
This collision process is implemented using Monte Carlo method and
the effect of Pauli exclusion principle is duly taken into account.
For the inelastic 
nucleon-nucleon channels we include the $\Delta(1232)$ as well as 
the $N^{*}(1440)$ resonance. In the intermediate energy range the 
resonance production is dominated by the $\Delta$, however, the 
$N^{*}$ yet gives non-negligible contributions to the high energetic 
pion yield \cite{metag93}. The 
resonances as well as the pions originating from their decay are 
explicitly treated, i.e. in a non-perturbative way and all relevant 
channels are taken into account. 
In particular we include the resonance production and rescattering 
by inelastic NN collisions, the one-pion decay of $\Delta$ and 
$N^{*}$ and the two-pion decay of the $N^{*}$ and one-pion 
reabsorption processes. ( For details see Ref. \cite{fuch96}.)

Pions thus produced are propagated according to the same equations
of motion given in Eq.( \ref{motion1}).
The Hamiltonian corresponding to pions is given as
\beq
H_\pi 
      = \sum_{i=1}^{N_\pi} \left( \sqrt{{\vec p}_{i}^2 + m_{\pi}^2} + 
 V^{\rm Coul}_i \right ).
\label{hamPi}
\eeq
The Coulomb interaction $V^{\rm Coul}_i$ between the baryons 
the charged pions is given as
\beq
V^{\rm Coul}_i = \sum_{j=1}^{N_b} \ {e_i e_j \over 
                               { \mid {\vec q}_j-{\vec q}_i \mid }}\quad ,
\label{picoul}
\eeq
where $N_b$ and $N_{\pi }$ are respectively 
the total number of baryons including charged resonances and pions
 at a given time $t$.
Thus, pion propagation at the intermediate and final stages is
guided essentially by the Coulomb effect.
\section{ Results and Discussions }
\subsection { High energy pions  and role of $N^*$ resonances}
Since the observation that high energy pions freeze out early when
emitted perpendicular to the reaction plane, the focus has been to
establish, both experimentally and theoretically
 that these high 
energy pions probe the hot and dense phase of the collision.
Hence, we analyse the properties of high energy pions and the
effect of heavier baryon resonances, such as the $N^*(1440)$, 
in this regard. Here we adopt the phenomenological Skyrme 
interaction SMD.

For this purpose we calculated the total pion multiplicity in 
Au+Au collision at 1 GeV/A corresponding to the {\it minimum
bias } condition. 
According to Ref. \cite{pelt96}, we choose $b_{\rm max}$=11 fm.
Results obtained are shown in Fig.1 as a function
of pion transverse momentum $p_t$ and pion freeze out time $t_{\rm freeze}$, 
i.e. the time after which pions do no more interact via inelastic 
scattering processes.
It can be seen that most of the pions are in the region $p_t \le 200$ MeV/c 
and $t_{\rm freeze} \sim 30$ fm/c.
In order to project out the high energy pions, we normalised the pion 
multiplicity $M(p_t)$ obtained for each value of $p_t$ by its maximum
value $M_{\rm max}(p_t)$. The normalized multiplicity is shown in Fig.2, 
where it is seen that the peak value shifts toward early times as the
value of $p_t$ is increased. For example, pions with $p_t \sim 200$ MeV/c
freeze out at $t \sim 30$ fm/c; whereas for $p_t \sim 500$ MeV/c one
finds $t_{\rm freeze} \sim 20$ fm/c. This directly shows that high$-p_t$ pions
originate from early times. In order to relate these pions with the
compressed early phase, we calculated the total baryon density $\rho_B$ in a
sphere of 2 fm radius around the collision centre with impact parameter 
b=1 fm. The time evolution of the $\rho_B/\rho_0$ is shown in Fig.3, 
where the $\rho_0=0.16\ fm^{-3}$ is the normal nuclear matter density.
The maximum density is achieved at about 25 fm/c and is about $2\rho_0$.
Over the time span $10 \le t \le 30$ fm/c the density in the collision
center is at least $\rho_0$, and this time span may then be identified with
the hot, compressed participant phase.
Thus, from this analysis we find similar to a previous 
study using hard equation of state and only $\Delta-$resonances \cite{bass94}
 that high energy pions with
$t_{\rm freeze} \sim 20$ fm/c pertain to the early phase of the collision,   
when the matter is compressed to about 2$\rho_0$.

For the sake of completeness we show in Fig.4 for two impact parameters, 
$b=0\ fm$ and $b=6\ fm$ the collision configuration, i.e. the
evolution of the average distance $\Delta R = R_P -R_T $ between the
centres of the colliding nuclei where $R_P$ and $R_T$ represent
the average position of the centres of the projectile and target
nuclei, respectively. The collision centre is at the zero of the
Y-axis. The curves representing $R_P$ are in the positive
Y-axis plane and those representing $R_T$ are in the negative
Y-axis plane. The two dotted lines correspond to the sharp density
radii of the two colliding nuclei.
Thereby, it can be seen that at about $4\ fm/c$ when the curves
corresponding to $R_P$ and $R_T$ intersect the dotted lines 
the nuclei come in contact, i.e. this represents 
the touching configuration. Further, 
for very central collisions the time
span upto $\sim 25 \ fm/c$ pertains to the compression stage, and
beyond which the expansion process begins.
Hence, the maximum density or compression is attained at about $25\ fm/c$, 
which is consistent with the results shown in Fig.3.
It may also be noted that for finite values of impact parameter $b$, 
one has at the time of maximum compression $\Delta R = b$.

To illustrate the contribution of high energy pions from $N^*$
resonances, we have compared in Fig.5 
the total pions originating from $\Delta-$
resonance to those from $N^*$. The pion multiplicity linearly weighted
by impact parameter b is shown as a function of b. 
It is seen that the contribution
from $N^*$--resonances irrespective of the nature of the collision, i.e. 
central, semi-central or peripheral, is about 5\%. Though their
multiplicity is small, they play an important role as most of them
originate from early times. This aspect is illustrated in Fig.6 where
we have plotted $M_{\pi}^{\Delta}/M_{\pi }$ as a function of pion freeze out
time $t_{\rm freeze}$. The total number of
 pions from $\Delta $ and $N^*$ are labelled
as $M_{\pi}^{\Delta}$ and $M_{\pi}^{N^*}$ respectively, with $M_{\pi} =
M_{\pi}^{\Delta}+M_{\pi}^{N^*}$.
Interestingly, the maximum contribution from $N^*$ (about $8\%$) occurs
at $t_{\rm freeze} \sim 20$ fm/c. It maybe recalled 
here that pions with this value of
$t_{\rm freeze}$ pertain to the dense phase with $\rho_B \ge 2\rho_0$.
Therefore, the majority of the $N^*$ pions originate 
from early times and hence
are most likely high energetic ones. 

To demonstrate this aspect we calculated the neutral pion spectrum at
midrapidity under the {\it minimum bias} condition. ( We have chosen
$\pi^0$ so that the high energy tail of the spectrum is not distorted
by Coulomb effects.)
Results obtained are shown in Fig.7 alongwith the 
$\pi^0$ spectrum calculated only with $N^*$ pions.
In the same figure, we have also shown the TAPS data
for comparision. Due to a recent experimental analysis the data 
given in \cite{schw94} have to be renormalized 
by factor 0.6 \cite{teis96}. 
It is clearly seen that the $N^*$ contribution to the high energy
tail is by a factor of about two more than its contribution toward low
energy pions. 
This is in agreement with a recent study \cite{teis96}
 based on the BUU model where it was shown, though for a 
different system at a higher energy, that
contributions from baryonic resonances like $N^*(1440,1520,1535)$ and
$\Delta (1600)$ are significant over the high energy tail of the pion
spectra.
Moreover, the $\pi^0(N^*)$ spectra in Fig.7 portrays two distinct
slopes and the slope parameter $T$ extracted from the high energy
part is relatively higher than the one obtained from the high energy 
part of the total $\pi^0$ spectrum.
This again establishes that $N^*$ pions pertain to earlier times as compared
to those from $\Delta $. Therefore, it may be said that
pions from $N^*$ need to be included for a consistent description
of the high energy tail.
\subsection {In-medium effects on charged pions }
As discussed in Introduction, we now investigate the effect 
of microscopically determined nuclear mean field on obsevables
such as pion yield and spectra.
(All data shown in this section are taken from
Ref. \cite{pelt96}. )

For this purpose, the G-matrix is obtained 
by solving the Bethe-Goldstone equation  
\begin{equation}
G = V + V {Q\over \omega-H_0 +i\eta }G
\end{equation}
where $Q$ is the Pauli operator which forbids nucleons from
scattering into occupied states. The bare NN interaction $V$
is taken to be the Reid soft-core interaction and $H_0$ 
is the single particle energy \cite{izum80,oths87}. Thus, 
the G-matrix accounts for the most relevant medium effects 
in nuclear matter. The mean field potential 
$U_{\rm GMAT}$ is derived from the G-matrix as \cite{jaen92,khoa92}
\begin{equation}
U_{\rm GMAT}(\rho,{\vec p}) = 4 \int_F {d^3p^{\prime}\over {(2\pi)}^3}
			({\vec p},{\vec p^{\prime}}
				\mid G(\rho) \mid
			 {\vec p},{\vec p^{\prime}}) 
\quad .
\label{ugmat}
\end{equation}
For the purpose of a simple application within the QMD 
model the potential is parametrized in a Skyrme-like form  
\begin{equation}
U_{\rm GMAT}(\rho,p) = a_1 {({\rho\over \rho_o})} +
a_2 {({\rho\over \rho_o})}^{a_3} +
a_4 ln^2[a_5p^2 +1 ] {({\rho\over \rho_o})}^{a_6} \quad .
\label{ufit}
\end{equation}
The coefficients $a_i's$ were determined from a fit of
Eq. (\ref{ufit}) to the microscopically calculated 
values of $U_{\rm GMAT}$, Eq.(\ref{ugmat}), within the density 
range of $0 \le \rho \le 2\rho_o$ and for momenta $p \le 500\ MeV/c$.
The momentum dependence of so-determined potential 
at $\rho = \rho_o$ is shown in Fig.8 alongwith the
one obtained from the phenomenological Skyrme force
using Eq. (\ref{skyrme}).
(Calculations done with the Skyrme force and the
G-matrix potential are hereafter referred to as
SMD and GMAT, respectively.)
For comparison, we have also shown the potential
as obtained by Li and Machleidt \cite{li93} using the 
Dirac-Brueckner approach.
It can be seen that the nuclear mean field derived
from the non-relativistic G-matrix is slightly 
more attractive at low momenta and less repulsive
at high momenta compared to the one obtained 
from the relativistic G-matrix. However, the general
behaviour is quite similar.
The prominent feature of the realistic G-matrix 
interaction is the significantly stronger 
momentum dependence as compared to the
Skyrme force. In a recent study \cite{fuchs96b} 
on the effect of realistic forces on collective
nucleon flow it was found that the relativistic
 (Dirac-Brueckner) and non-relativistic (Brueckner)
approaches yield quite similar results when applied 
to heavy ion collisions. However, 
the differences to phenomenological interactions 
(Skyrme or Walecka model) are significant.
Thus, it is of interest to investigate the sensitivity
of pionic observables to realistic forces.
In order to be as consistent as possible, we have 
also considered the effect of in-medium 
cross sections \cite{jaen92,khoa92} in the
elastic NN channel for center-of-mass 
momentum $p < 500\ MeV/c$, which are
derived from the same microscopic 
G-matrix potential using the two-Fermi-sphere 
aprroach. However, the effect of the in-medium 
cross section on the pion yield has been found to 
ba small compared to that of the mean field and is 
therefore not discussed in detail in the following.

Results obtained for the $\pi^+$ multiplicity
is plotted in Fig.9 as a function of the 
participanting nucleons $A_{\rm part}$. 
It is seen that the theoretical calculation
overestimate the data of the $Au$ system by
about $50\%$ at small impact parameters.
In this context we want to mention that 
the pion multiplicities obtained by other
groups for central $Au + Au$ reaction at
1 GeV/A, using QMD \cite{bass95,pelt96}
or BUU \cite{teis97,helg97} models, 
 differ only within about $10\%$ with our results.
Thus, this overestimation at small
impact parameters seems to be a general
feature of present transport calculations. However, the theoretical
prediction is more in agreement with the 
experimentally derived Harris systematics \cite{harr87}. 
It may also be noted that the Harris
systematics while it overpredicts as compared
to the FOPI data in the case of heavy system such as $Au$ 
is in good agreement with the data 
for smaller system like $Ni$ \cite{pelt97}.
 We have also shown in Fig.10, the rapidity distribution obtained 
under minimum bias condition. In both SMD
and GMAT calculations one finds at midrapidity
a overestimation of about $30\%$, as compared
to the data. Similar features are also observed
in the case of $\pi^-$. 

To see the effect upon momentum distribution, 
we calculated the charged pion spectra. 
Results obtained with both SMD and GMAT potentials
are then compared in Figs.11 and 12 to the 
corresponding FOPI data. The overall agreement 
with the observed spectra is reasonably good.
In particular, one finds that with the inclusion
of in-medium effects, there is a slight reduction
over the low energy part thus coming
closer to the data. But, a larger effect
is observed over the high energy part where the
subthreshold pions get quite suppressed indicating
the importance of resonances heavier than $\Delta$
and $N^*$. Thus, one finds that, as compared to the high 
energy part, the low energy part is not too much
affected by the strong momentum dependence of
GMAT potential which may be due to the fact 
that low energy pions are frequently reabsorbed.
For a more complete description of pion yields
and spectra, one probably needs to include the
in-medium effects in the inelastic NN channel
as well. In addition, a possible reason for the 
overestimation over the low energy part
can be the life-time of the resonances
involved. ( Some indication towards this can be inferred
from Fig.1, where one finds that even at $60\ fm/c$, 
resonances are still decaying into pions though
their multiplicity is quite small. )
E.g. in Ref.\cite{dani96} an alternative treatment of the
resonance lifetime has been proposed, which is based on
more quantum mechanical aspects. However, it is still
unclear at the moment how such a description should be
applied consistently in transport simulations.
\subsection {$\pi^-/\pi^+$ ratio and Coulomb effect }
We had demonstrated theoretically that high energy pions originate
from the early phase of the collision.
Here, using the observed energy dependence of $\pi^-/\pi^+$ ratio, 
we re-iterate
that high energy pions do probe the hot and dense matter.

It is known that the two principal effects contributing toward the
observed charged dependence of pion production, i.e.
$\sigma (\pi^-) > \sigma (\pi^0 ) > \sigma (\pi^+) $, are the isospin
dependence of the pion production cross sections and the Coulomb
force. As most of the pions observed are from $\Delta-$resonance, 
one can determine the effect of isospin. Using the isobar model 
\cite{stoc86} one finds for pion originating from $\Delta$--decay 
the general relation
\beqa 
\sigma (\pi^- )/\sigma (\pi^0 ) = { 5N^2+NZ\over N^2+4NZ+Z^2 } 
\quad, \quad
\sigma (\pi^+ )/\sigma (\pi^0 ) = { 5Z^2+NZ\over N^2+4NZ+Z^2 } 
\quad . 
\label{isobar}
\eeqa 
For the Au+Au system we obtain the ratios $\pi^-:\pi^0:\pi^+ \ = \ 
1.37:1:0.71$ for the repective multiplicities. 
Therefore, one has $\pi^-/\pi^+ = 1.95$. 
For pions from $N^*$ the isobar model yields 
$\pi^-/\pi^+ = 1.70$. 
Since these relations are globally valid, i.e. independent of the
pion energy, the observed energy dependence of 
the $\pi^-/\pi^+$ ratio must be due to the 
Coulomb potential energy $V_C$ \cite{bass94,teis97}.
This is illustrated in Fig. 13, where the ratio 
of the calculated differential cross sections 
$\sigma_E (\pi^- )/\sigma_E (\pi^+ )$ with 
$\sigma_E = \d^2\sigma/(pE\d p\d\Omega )$
are compared with the observed one \cite{pelt96}. 
One finds that the overall behaviour of the measured ratios 
is well reproduced. 
In particular, over the low energy part both SMD and GMAT
curves are quite similar. However, at higher energies
$T_{\rm c.m.} > 300$ MeV, i.e. $p_{\rm c.m.} > 400$ MeV, 
the GMAT results show stronger fluctuation as compared to
the SMD case, which may be due to the suppressed 
cross sections.
Such strong fluctuations over high energy part are also 
apparent in the experimental data. Nevertheless, the
trend in Fig.13 indicates that the results approach a constant 
value as the pion momentum $p_{\rm c.m.}$ increases which 
is in agreement with the experimental findings. 

The general features of the 
effect of the Coulomb force on the $\pi^-/\pi^+$ can be understood
using a simple analytical model as described in Ref. \cite{gyul81}. 
Let ${ p_0 }$ and ${ p }$ be the initial and
final momenta of a charged particle. Then the charged particle
cross section $\sigma ({  p}) \equiv \d^3\sigma/d^3p$ can
be related to the neutral particle cross section
$\sigma_0 ({ p}) \equiv \d^3\sigma_0/ \d^3p$ through the
change of 
\beq
\sigma( {  p } ) = \sigma_0 ( { p_0 }({ p} ) ) \mid {
\frac{\partial^3p_0}{\partial^3p} }\mid.
\label{coulcrsec}
\eeq
The two effects due to Coulomb force are the momentum shift 
$\delta { p} = { p} - { p_0}$ and the Coulomb phase
space factor $\mid {\partial^3p_0/\partial^3p }\mid $.
Taking relativistic quantum corrections up to first order
in $Z\alpha $ and in the limit ${ p} \longrightarrow \infty $, one
has $\pm \delta p  \sim V_C/\beta_p $ and 
$\mid {
\partial^3p_0/\partial^3p }\mid \sim 1 \pm V_C/p$, where
$\beta_p = p/\omega_p$ and $\omega_p = \sqrt { p^2 + m_{\pi }^2 }$.
 Omitting the phase space factor in Eq. ({\ref{coulcrsec}})
in the limit ${ p} \longrightarrow \infty $ 
the $\pi^-/\pi^+ $ ratio is then given as
\beq
{\sigma_{\pi^-}\over \sigma_{\pi^+}} \sim {C_{-}\over C_{+}}
\ exp \left ( -{1\over T}[\omega_{-}-\omega_{+}] \right )
\label{ratio}
\eeq
where $\omega_{\pm}^2 = { p}_{\pm }^2 + m_{\pi }^2 $ and
${ p}_{\pm} = { p} \mp \delta { p}$.

It is found experimentally that the
high energy slopes of $\pi^-$ and $\pi^+$ are very similar and
hence the slope parameter $T$ is the same for both positive 
and negative charged
pions. Moreover, one finds Eq. ({\ref{ratio}}) is independent
of the momentum $p$ and the ratio $\pi^- / \pi^+$ is in good 
approximation given by a constant determined by the Coulomb energy 
shift, i.e. by $V_C$. This is consistent 
with the experimental observation where one finds that the
$\pi^-/\pi^+$ ratio approaches a constant value of about 0.93 
in the high momentum limit. Therefore, the above approximate equation
can be used in the high energy limit 
to estimate the Coulomb energy $V_C$. 
In addition we take the value of $C_{-}/C_{+}$ 
according to Eq. {\ref{isobar}} to be 1.95 thus accounting for the 
isospin dependence of the cross sections. It may be noted that 
a very similar value $(C_{-}/C_{+} = 1.94\pm0.05) $ is obtained 
experimentally in the zero momentum limit \cite{oesc96}. 
Taking the slope parameter $T$ to be 94.6 MeV \cite{pelt96} 
and the constant value of
$\pi^-/\pi^+$ ratio to be 0.93 we obtain the Coulomb energy $V_C$
to be about 35 MeV. The radius $R$ of the pion emitting source is then
estimated to be 4.5 fm with the assumption of a static uniform 
Coulomb field $V_C \sim Ze^2/R$ where
the total number of participating charges $Z$ is taken to be 
110 as estimated in Ref. \cite{oesc96}.
We also estimated the density $\rho \sim 3A/(4\pi R^3)$ 
of the participant zone to be about $4\rho_0 $ where the
total number of participants $A \simeq 197\ (Z/79)$.
It may be noted that the estimation of $\rho $ is about two times the value
obtained from the actual simulation, see also Fig. 3. 
If we consider a more realistic
value $\rho = 2\rho_0$ and using the fact that  
$V_C \simeq 35$ MeV we obtain $R=6.7$.
This then gives an idea about the order of magnitude of
uncertainity present in the estimation of the pion source size $R$
and the effective number of participating charges $Z$. 
From this simple analysis based on the observed energy
dependence of the $\pi^-/\pi^+$ ratio, we find in support of
 our theoretical investigation that the high energy pions 
probe the hot, compressed phase of the collision. However, 
the empirical determination of the freeze-out radius from measured 
$\pi^-/\pi^+$ ratios \cite{oesc96} should be performed carefully. 
In this context, it may be noted that a recent analysis \cite {osad96} of 
$\pi^-/\pi^+$ ratios measured in $Au\ + \ Au$ collision at
11 GeV/A and in $Pb\ + \ Pb$ at 158 GeV/A shows valley like
structures in the minimum $\chi^2$ plane rendering the
determination of $R$ and $Z$ from the data difficult.
Moreover, the scenario for the origin of high energy pions is 
probably not that of an equilibrated and thermalized source but 
rather that of two counterstreaming nuclear matter currents with 
high relative velocities and thus, e.g., Lorentz forces should 
be taken into account in such an analysis. 
\section {Summary}
In this work we investigated the production
of pions in Au+Au collisions at 1 GeV/A using
the QMD model. Focussing 
upon the high energy pions, we showed that these subthreshold 
pions freeze out at early times when the matter is compressed 
to more than two times the saturation density. 
Hence, they probe the hot and dense phase of the
nuclear reaction. One of the processes which can produce such
energetic pions is the decay of baryonic resonances like the $N^*$.
A detailed analysis shows that eventhough the multiplicity of pions
from the decay of $N^*$ is small compared to those from $\Delta-$decay, 
i.e. about 5$\%$ of the total yield, 
they play an important role as most of them pertain to the early phase
of the reaction. Further, it is found that $N^*$ contribution toward
high energy pions is by a factor of about two more than its contribution
toward low energy pions. Therefore, pions from higher resonances like $N^*$
need to be included for a consistent description of the high energy tail
of the pion spectra.

Moreover, due to the importance of momentum dependence in the
study of heavy ion collisions, we have also studied the role
of in-medium effects by performing calculations with 
a microscopic nuclear mean field derived from the
Brueckner G-matrix using the Reid soft-core interaction. 
It is found that calculated pion multiplicities, 
though they agree reasonably
well with the Harris systematics overestimate the FOPI data
at small impact parameters by about $50\%$.
The use of realistic G-matrix interaction improves this
situation, however, not to a sufficiently high extent.
 We then compared the
charged pion spectra calculated with G-matrix and Skyrme 
potentials to the respective FOPI data. It is found that
the high energy part gets relatively more suppressed by
in-medium effects, as compared to the low energy part.
Therefore, for a complete description of pion yield and 
spectra, one probably needs to consistently include in-medium 
effects both in the elastic and inelastic NN channels.
In addition, an improved treatment of lifetime of
resonances may also be warranted.

Finally, we investigated the isospin dependence of the pion 
production. The present calculations are able to reproduce the 
overall behavior of the measured $\pi^-/\pi^+$ ratios. 
Deviations from the simple isobar model originate from the 
Coulomb force. These deviations found for 
high energy pions indicate the influence of the Coulomb 
field on pions stemming from the high density 
phase. These findings are in qualitative agreement with a simple 
analysis based only on the observed $\pi^-/\pi^+$ ratios and the slopes 
of the respective spectra. \\
\begin{ack}
We thank D. Pelte for providing us
with the FOPI data shown in this work.
\end{ack}
\newpage

\newpage
\newpage
\begin{figure}
\begin{center}
\leavevmode
\epsfxsize = 15cm
\epsffile[0 0 450 420]{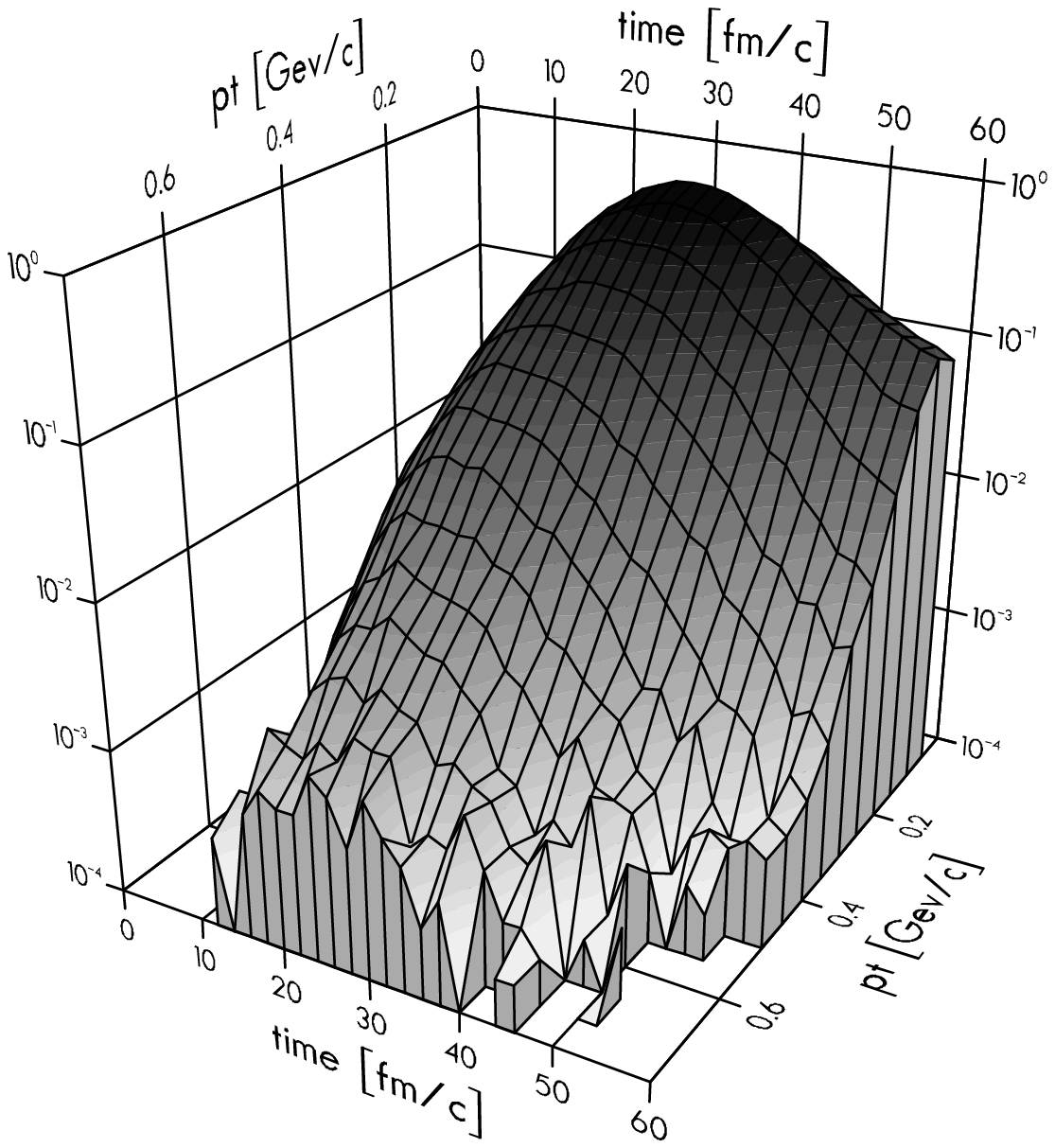}
\end{center}
\caption{
Total pion multiplicity calculated with a Skyrme force (SMD) 
in Au+Au at 1 GeV/A under
{\it minimum bias } condition is shown as a function of pion
transverse momentum ${ p}_t$ and pion freeze out time $t_{\rm freeze}$.
}
\label{fig1}
\end{figure}
\begin{figure}
\begin{center}
\leavevmode
\epsfxsize = 15cm
\epsffile[0 0 450 420]{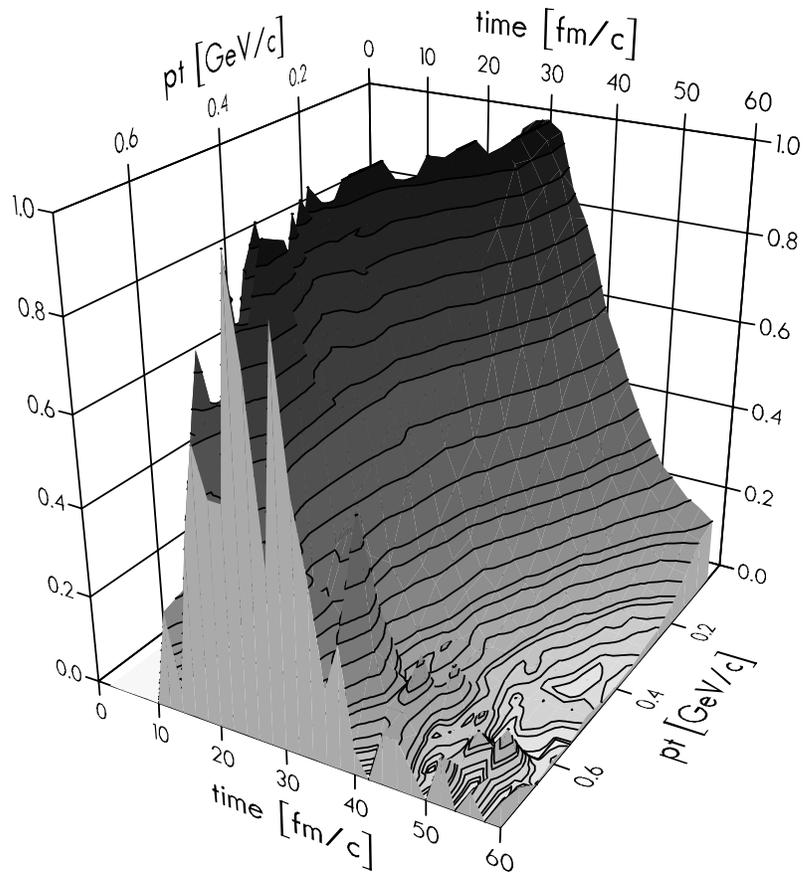}
\end{center}
\caption{
Normalized pion multiplicity, i.e. the 
multiplicity calculated for each value of ${ p}_t$ 
is normalized by its maximum value, 
obtained in as the same reaction in Fig. 1
is shown as a function of pion
transverse momentum ${{ p }}_t$ and pion freeze out time $t_{\rm freeze}$.
}
\label{fig2}
\end{figure}
\begin{figure}
\begin{center}
\leavevmode
\epsfxsize = 13cm
\epsffile[0 60 520 510]{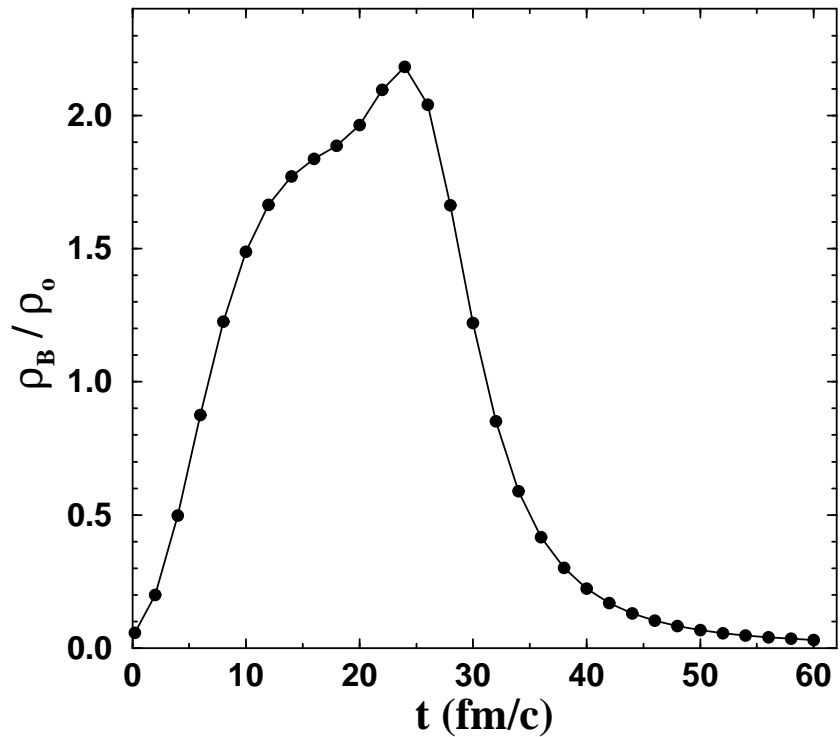}
\end{center}
\caption{
Time evolution of the baryon density $\rho_B$ in a central ( b=1 fm) 
Au on Au reaction at 1 GeV/A. $\rho_B$ is calculated 
with a Skyrme force (SMD) in a 
sphere of 2 fm radius around the collision center. 
}
\label{fig3}
\end{figure}
\begin{figure}
\begin{center}
\leavevmode
\epsfxsize = 13cm
\epsffile[0 60 550 510]{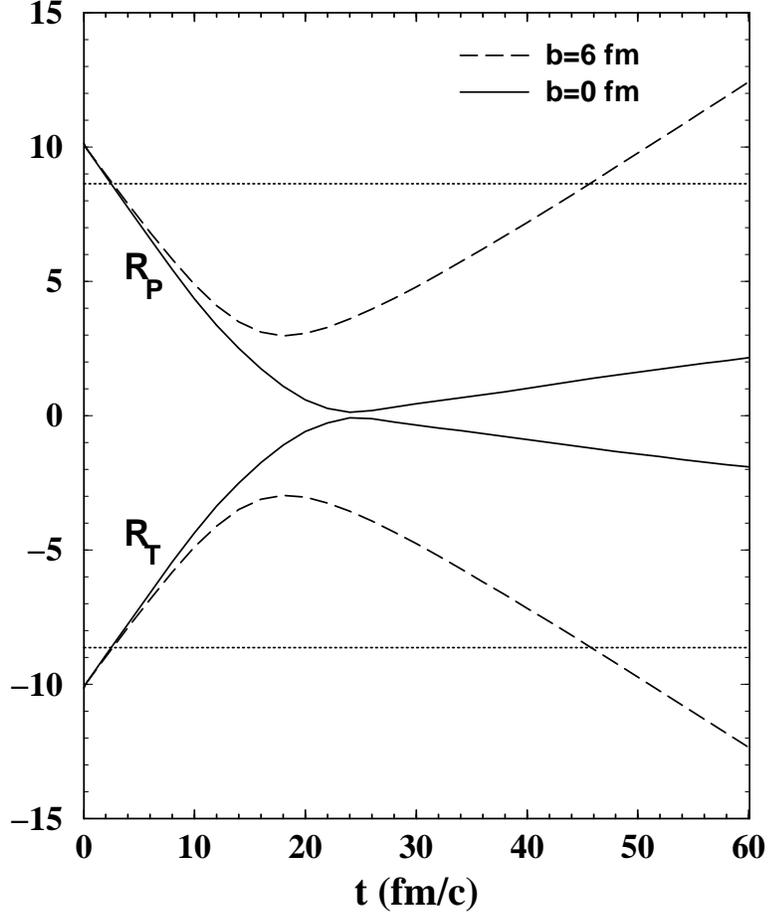}
\end{center}
\caption{
Time evolution of the average distance $\Delta R=R_P-R_T$ 
between the centres of the two
colliding nuclei ($Au\ + \ Au$) is shown for two impact parameters. 
$R_P$ and $R_T$ represent the average position of the 
centre of the projectile
and target, respectively.
Curves corresponding to $R_P$ are in the positive Y-axis plane, and 
those for $R_T$ are
in the negative Y-axis plane.
Zero of the ordinate 
corresponds to the collision centre. 
Dotted lines
represent the sharp density radii of the two nuclei. 
}
\label{fig4}
\end{figure}
\begin{figure}
\begin{center}
\leavevmode
\epsfxsize = 13cm
\epsffile[0 60 550 510]{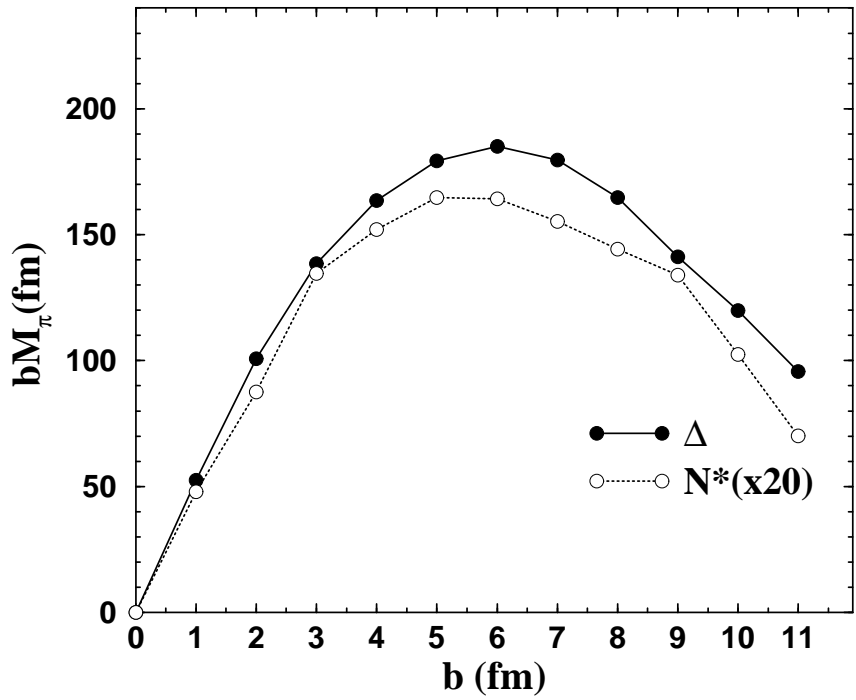}
\end{center}
\caption{
The total multiplicity of pions calculated with a Skyrme 
force (SMD) originating from $N^*$--resonances 
is compared to those from $\Delta$--resonances.
The pion multiplicity $M_{\pi}$ linearly weighted by the impact parameter 
b is shown as a function of b. 
}
\label{fig5}
\end{figure}
\begin{figure}
\begin{center}
\leavevmode
\epsfxsize = 13cm
\epsffile[0 60 575 440]{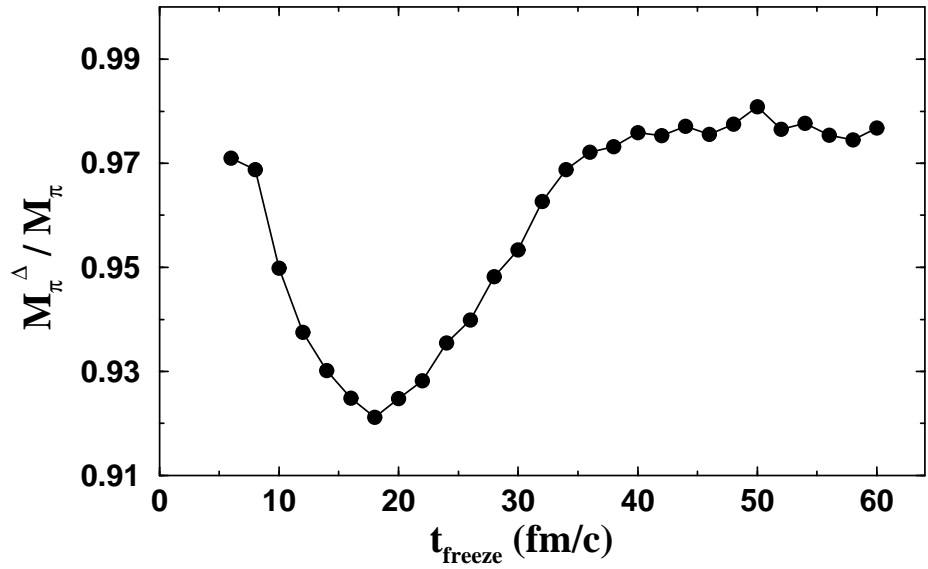}
\end{center}
\caption{
The multiplicity of pions stemming from the decay of 
$\Delta$--resonances $M_{\pi}^{\Delta}$ 
normalized to the total pion multiplicity $M_{\pi}$ calculated 
with a Skyrme force (SMD) is shown as a
function of the pion freeze-out time $t_{\rm freeze}$. 
}
\label{fig6}
\end{figure}
\begin{figure}
\begin{center}
\leavevmode
\epsfxsize = 13cm
\epsffile[0 60 530 480]{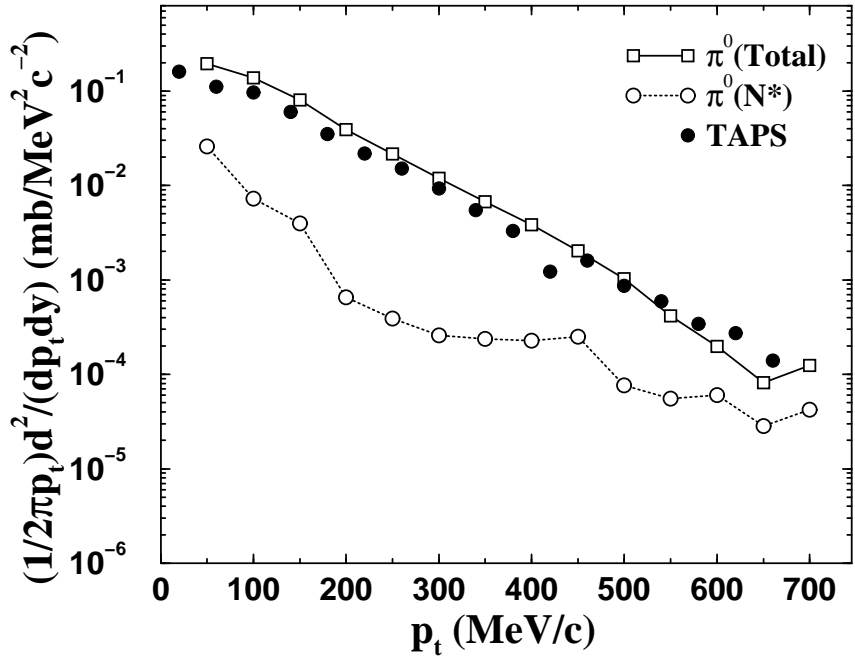}
\end{center}
\caption{
 Neutral pion $p_t$ spectrum calculated
 with a Skyrme force (SMD) 
at midrapidity for a Au on Au reaction at 1 GeV/A
under {\it minimum bias} condition is compared 
with the corresponding 
$\pi^0$ spectrum of pions originating from the decay of 
$N^*$--resonances. For comparison, TAPS data 
given in Ref. \protect\cite{schw94} is also shown.
Due to a recent experimental analysis the data are 
renormalized by factor 0.6.
}
\label{fig7}
\end{figure}
\begin{figure}
\begin{center}
\leavevmode
\epsfxsize = 13cm
\epsffile[0 60 530 480]{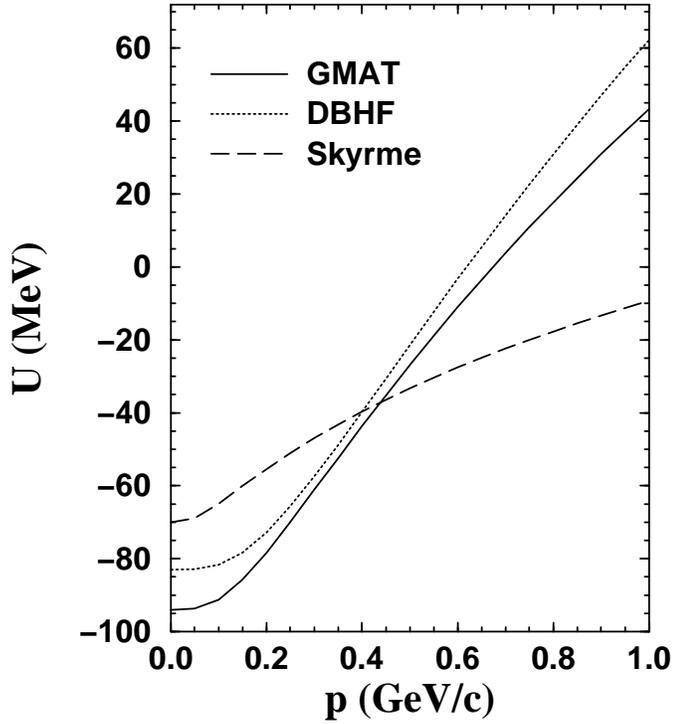}
\end{center}
\caption{
The momentum dependence of nuclear single particle potential 
derived from the Brueckner G-matrix (GMAT) using the realistic
Reid soft-core NN interaction is compared to the Skyrme 
interaction (SMD) at a density $\rho=\rho_o$. 
For comparison, also the potential obtained in Ref. \protect\cite{li93}
using the relativistic Dirac-Brueckner approach (DBHF) is shown.
}
\label{fig8}
\end{figure}
\begin{figure}
\begin{center}
\leavevmode
\epsfxsize = 13cm
\epsffile[0 60 530 480]{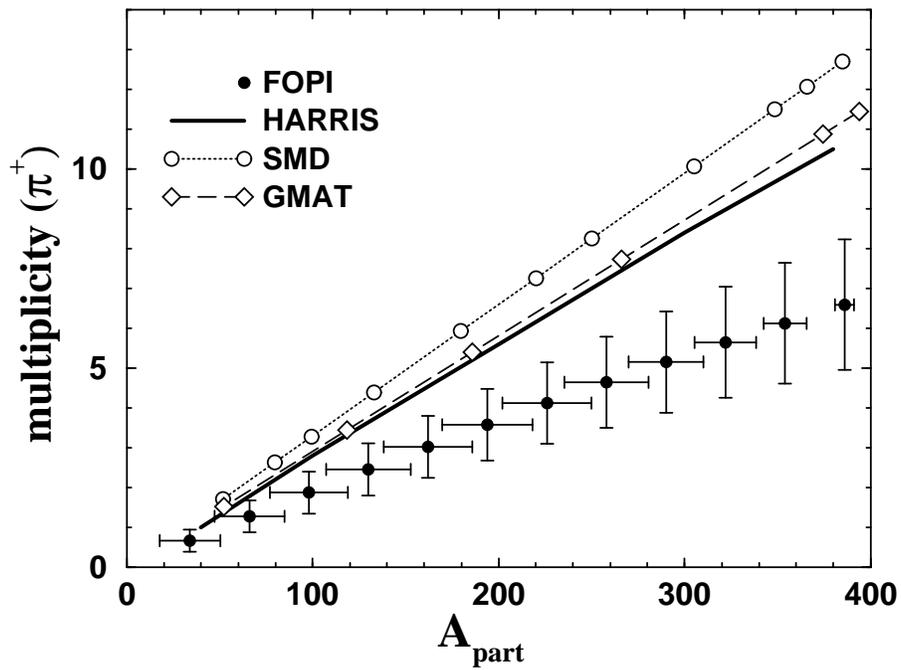}
\end{center}
\caption{
The participant number dependence of $\pi^+$ 
multiplicity as obtained from 
the Brueckner G-matrix (GMAT) and
a Skyrme force (SMD) are compared with the
FOPI data \protect\cite{pelt96}. In addition the 
experimentally derived Harris systematics 
\protect\cite{harr87} is shown.
}
\label{fig9}
\end{figure}
\begin{figure}
\begin{center}
\leavevmode
\epsfxsize = 13cm
\epsffile[0 60 530 480]{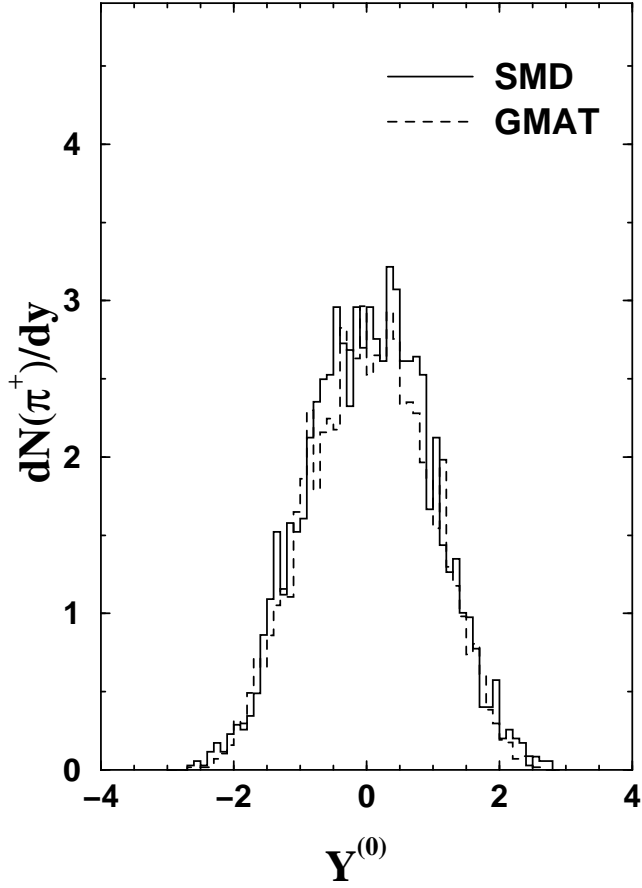}
\end{center}
\caption{
The rapidity distributions of $\pi^+$ 
 as obtained from 
the Brueckner G-matrix (GMAT) and a 
Skyrme force (SMD) is shown for transverse momenta
$p_t > 0.1\ GeV/c$. The scaled c.m. rapidity 
$Y^{(0)} = Y/Y_{\rm proj}$ is used.
}
\label{fig10}
\end{figure}
\begin{figure}
\begin{center}
\leavevmode
\epsfxsize = 13cm
\epsffile[0 60 530 480]{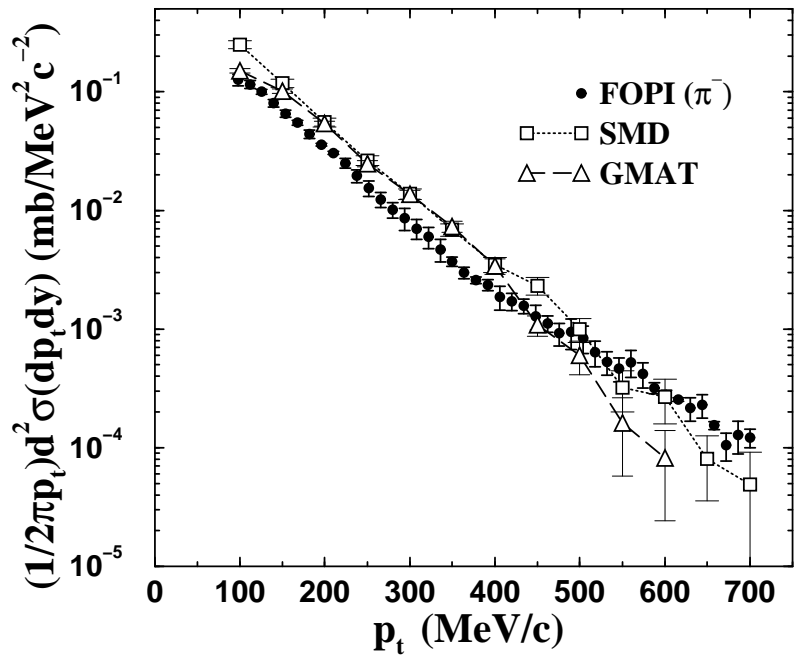}
\end{center}
\caption{
The $\pi^-$ $p_t$ spectrum obtained at midrapidity in a 
Au on Au reaction at 1 GeV/A
under {\it minimum bias} condition is compared to the 
FOPI data of Ref. \protect\cite{pelt96}.
}
\label{fig11}
\end{figure}
\clearpage
\begin{figure}
\begin{center}
\leavevmode
\epsfxsize = 13cm
\epsffile[0 60 510 430]{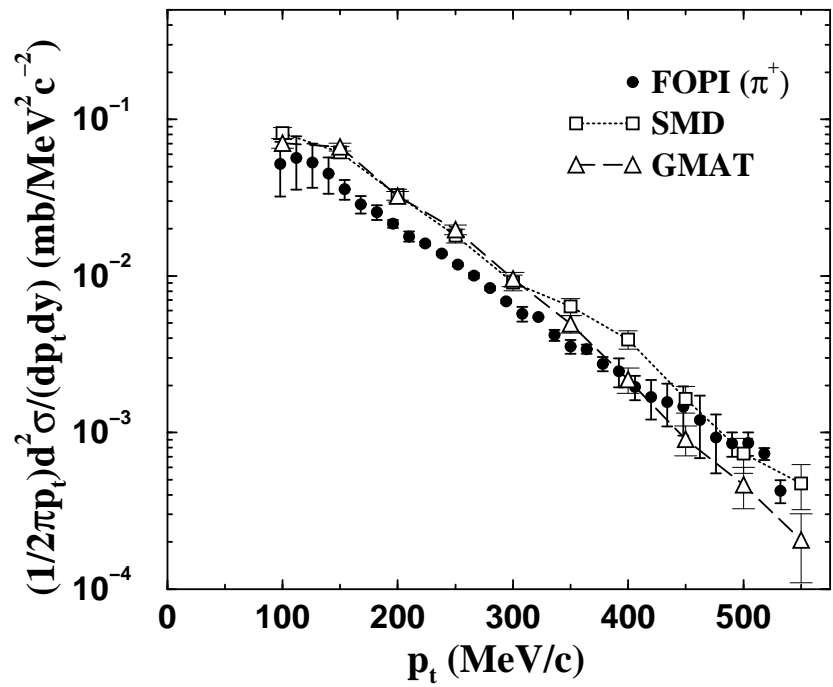}
\end{center}
\caption{
Same as Fig. 11, but for $\pi^+$.
}
\label{fig12}
\end{figure}
\clearpage
\begin{figure}
\begin{center}
\leavevmode
\epsfxsize = 13cm
\epsffile[0 60 510 430]{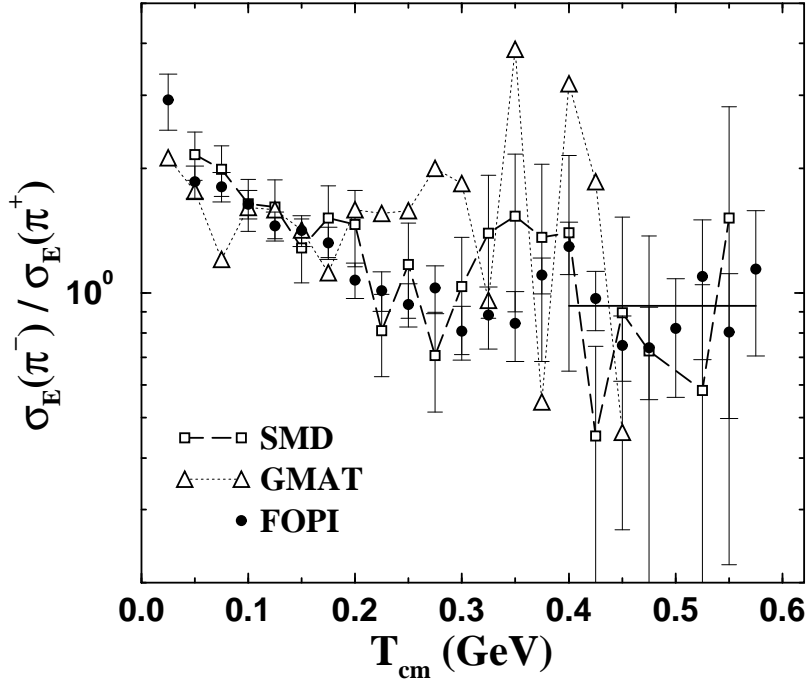}
\end{center}
\caption{
The $\sigma_E (\pi^- )/\sigma_E (\pi^+ )$ ratio calculated for the 
same reaction as in Fig.11. at midrapidity under 
{\it minimum bias } condition is shown as a function of the 
center-of-mass kinetic energy $T_{\rm c.m.}$. The corresponding 
cross sections $\sigma_E = d^2\sigma/(pEdEd\Omega )$ are 
obtianed with an angular cut $85^0 < \Theta_{\rm c.m.} < 135^0$. 
The data points are taken from the FOPI 
collaboration \protect\cite{pelt96}. 
The horizontal solid line indicates that the observed 
$\pi^-/\pi^+$ ratio reaches a constant value of about 0.93 as pion
energy increases.
}
\label{fig13}
\end{figure}
\end{document}